\documentclass[twocolumn,showpacs,showkeys,preprintnumbers,amsmath,amssymb]{revtex4}
\usepackage{graphicx}
\usepackage{dcolumn}
\usepackage{bm}
\begin{document}
\title{The Ramsauer model for the total cross sections of  neutron nucleus scattering}
\author{R. S. Gowda$^+$, S. S. V.   Suryanarayana$^\dagger$\footnote  {\normalsize{  \it  The
author Suryanarayan's name appeared  also as \\  S. V. S.  Sastry    in
Nuclear  physics  journals. }} and S. Ganesan$^+$}
\affiliation{  $^\dagger$  Nuclear  Physics Division, $^+$ Reactor Physics Design
Division,\\
Bhabha Atomic Research Centre, Trombay, Mumbai 400 085, India}
\begin{abstract}{
Theoretical  study  of  systematics of neutron scattering cross sections on
various materials for neutron energies up to several  hundred  MeV  are  of
practical  importance. In this paper, we analysed the experimental
neutron scattering total cross sections from 20MeV to 550MeV using Ramsauer model
for nuclei ranging from Be to Pb.
}\end{abstract}
\date  {\today}
\pacs{ 24.10.Ht, 25.40.-h, 28.20.Cz}
\keywords{total  cross  section, neutron nucleus scattering,
Ramsauer model,  empirical formulae.}
\maketitle
\par
\noindent
In recent times world over,  there is a renewed interest in the neutron nucleus scattering data.
This is owing to a new concept of controlled nuclear energy source called
the  accelerator driven sub-critical (ADS)
system \cite{ads1,ads2}.  In  this   ADS
system,  neutrons  are produced by bombarding a heavy element target with a
high energy proton beam of typically above 1.0GeV with a current of $>10mA$
\cite{ads1}. The sub-critical reactor is driven critical by spallation neutrons
produced by proton beam on typically a molten lead target.
Such accelerator driven systems can  also be used for   waste  incineration
of the long lived radio active waste produced in reactors based on thermal neutron induced fission.
Reactor physics calculations of these ADS type of systems
require neutron-nucleus scattering cross sections up to  500 MeV neutron energy.
Unlike the neutron energy spectrum from a thermal neutron fission, the spallation neutron
energies reach up to several hundred MeV.
Therefore, it is currently very important to  study the  systematics
of neutron scattering cross  sections  on  various nuclei  for  neutron
energies up to  several  hundred  MeV.  In the
present work, we performed an  empirical study of  the  energy
dependence  of  total  cross sections ($\sigma_{tot}$)
of the neutron-nucleus  (n-N)  scattering.  It is well known that the  total
cross  sections are explained  by  the nuclear Ramsauer model. The
Ramsauer model was first proposed by Lawson \cite{lawson} 
in the year 1953  to phenomenologically
explain  the  energy  dependence  of  total cross sections of
neutron nucleus scattering. In  order  to  have insight of the working of this model,  it  is
necessary  to understand the optical model (OM) description of neutron
scattering. In the OM approach, complex optical model potentials (OMP)  are
used  and  the  Schrodinger's  equation  is solved to obtain the scattering
amplitude. The real part of  the  OMP  describes  the  scattering  and  the
imaginary  part  results in attenuation or absorption of the incident wave.
The reaction cross section is given by the absorption of the neutron flux.
The  scattering calculations  are  performed  using  partial  wave
expansion   method   and   the   phase    shifts    ($\eta_\ell=\alpha_\ell
e^{i\beta_\ell})$  are  determined.  These complex phase sifts are strongly
angular momentum and energy dependent for a given  set  of  potentials.  In
terms  of  the  phase  shifts  and  the  wave  number ($ \lambdabar = \hbar
/\sqrt{2mE}$), the total cross sections are given below.
\begin{eqnarray}
\sigma_{tot}&=&2\pi\lambdabar^2\sum_\ell(2\ell+1)\left[1-\Re{\eta_\ell }\right] \label{st-om}
\end{eqnarray}
\noindent
Extensive  study  of the optical model fits of scattering cross sections on
various nuclei over wide energy range have been  made  by  several  groups.
This  is  owing  to  the  excellent  data  base  of  neutron total cross sections
available in the energy range up to 600 MeV  \cite{finlay,dietrich1,abfal}.
The  most  recent  work  by Koning and Delaroche (KD) \cite{kd} presents a very
exhaustive search for OMP parameters that fit the data very well up to  200
MeV.  Alternatively, the  nuclear
Ramsauer  model  \cite{lawson}  provides a simple means to parameterise the
energy dependence of neutron nucleus total scattering cross sections.  This
model  assumes  that the scattering phase shifts are independent of angular
momentum ($\ell$) as  given  in  Eq.(2)  ($\eta=\alpha  e^{i\beta})$  ,  in
contrast  to the predictions of the optical model given in Eq.\eqref{st-om}. Further,
it was proposed that the $\ell$-independent phase shift varies slowly  with
energy. This model was successfully  applied for  neutron  scattering
from  various  nuclei  by  Peterson  \cite{peterson,book}.  There were some
attempts \cite{franco,gould,anderson,grimes1}  (see references therein)
to  put  this  Ramsauer  model  on  a  sound  theoretical   basis. The neutron
total cross sections have thus been well studied using this model,  over  a
wide  range  of  nuclear  masses  as well as neutron energies up to 500 MeV
\cite{anderson,bauer,madsen,grimes1,grimes2,grimes3,dietrich2}.
Deb {\it et. al.,} \cite{deb} have achieved simple functional forms for the total cross sections
by parameterising the maximum partial wave ($\ell_0=\ell_{max}$) values.
\noindent
\par
In our earlier work \cite{surya1}, we  presented the Ramsauer model analysis of
the results of optical model  code SCAT2 \cite{scat2} using KD potentials.
In the present work, we  performed  the  Ramsauer  model  analysis  of the experimental data of
neutron total cross sections for heavy  and  light nuclei by using
Eq.\eqref{eqstfit}.  The quantities
$R (fm),\alpha,\beta$ are functions of atomic mass number (A)  and the center of mass
energy (E).
\begin{eqnarray}
\sigma_{tot}&=&{2\pi}\ (R+\lambdabar)^2\left(1-\alpha\cos{\beta}\right) \label{eqstfit} \\
\beta&=&\beta_x A^{\frac{1}{3}}( \sqrt{E+V} - \sqrt{E} )\label{stfitfun}\\
V&=&V_A+V_E\sqrt{E}\nonumber  \\
V_A&=&V_0+V_1 (N-Z)/A +V_2/A \nonumber\\
\alpha&=&\alpha_0+\alpha_A \sqrt{E} \nonumber\\
\alpha_A&=&\alpha_{1} \ln(A) +\alpha_{2}/\ln(A)  \nonumber\\
R&=&r_0 A^{\frac{1}{3}} + r_A \sqrt{E} + r_2  \nonumber\\
r_A&=&r_{10} \ln(A) + r_{11}/\ln(A) \nonumber \\
r_0&=&1.42988,~~r_{10}=-0.02298,~~r_{11}=0.10268 \label{stfitpar}\\
r_2&=&0.23216,~~V_0=46.51099,~~V_1=6.73833 \nonumber\\
V_2&=&-117.52082,~~V_E=-3.21817,~~\beta_x=0.5928 \nonumber\\
\alpha_0&=&0.02868,~~\alpha_{1}=-0.00274,~~\alpha_{2}=0.13211 \nonumber
\end{eqnarray}
Figure 1 shows the Ramsauer model fits (solid lines) for $\sigma_{tot}$  cross  sections
using Eqs.\eqref{eqstfit},\eqref{stfitfun},\eqref{stfitpar} and the symbols  represent
the experimental data. The fits are
obtained with total of twelve free parameters as given in   Eq.\eqref{stfitpar},  over  wide
mass  range of $^{24}Mg$ to $^{208}Pb$. In Fig.2, we show the Ramsauer model fits for light nuclei
from $^9$Be to $^{24}$Mg.
These fits   cover the neutron energy region ($E_{cm}$) of 20-550 MeV.
Similar Ramsauer model fits to total cross
sections were already shown by various groups  \cite{grimes1,grimes2,deb}  (see  the  references  therein).
As shown in Figs. (1,2), the  functional  dependence  on  energy  and  mass given in Eqs.\eqref{eqstfit},\eqref{stfitfun}
with twelve global parameters was able to reproduce the experimental data very  well.
\begin{figure}
\includegraphics[width=8.5cm,height=9.50cm]{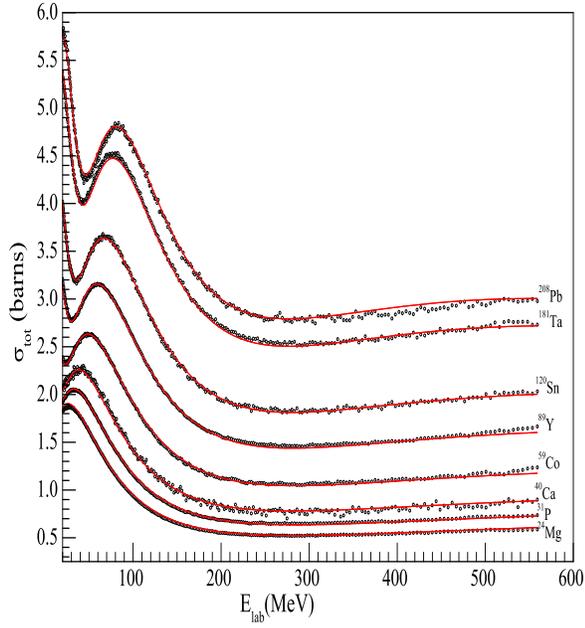}
\caption{Ramsauer  model  fits  (solid  lines)  to  experimental neutron total  cross
sections (symbols) versus  E$_{lab}$,  using  Eq.\eqref{eqstfit}.
The  twelve parameters required are mentioned in Eq.\eqref{stfitpar}.
The fits are for heavy nuclei as shown by the figure labels. }
\label{sigtheavy}
\end{figure}
\begin{figure}
\includegraphics[width=8.5cm,height=9.50cm]{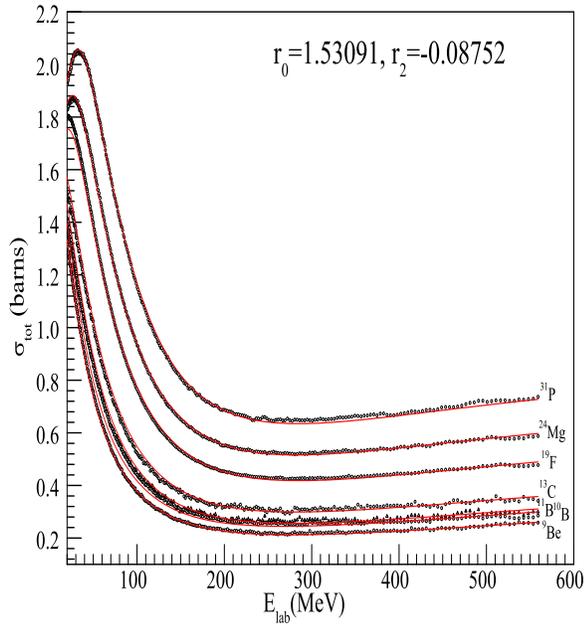}
\caption{Ramsauer  model  fits  (solid  lines) to light nuclei, similar to Fig. 1.
The  twelve parameters required are same as in Fig. 1, except that the
radius parameters for light systems are different as mentioned on figure.}
\label{sigtlight}
\end{figure}
As shown in Fig. 2, the parameters are same as those for heavy systems, except for the radius parameter.
The radius parameter for scattering from light nuclei turns out to be rather larger than for heavy nuclei.
In some cases of neutron nucleus scattering, the radius parameter (only r$_0$)  has to be specially adjusted to achieve
best fits. These cases did not obey the  Ramsauer model systematics for either light or heavy nuclei.
In case of $^{27}$Al, $^{232}$Th and $^{238}$U,   the adjusted radius parameters (in fm) are respectively,
r$_0$= 1.3892(Al), 1.4488(Th), 1.44641(U). These values are comparable to the general systematics of  r$_0$ values
for heavy nuclei (1.42988fm) and for light nuclei (1.53091fm).
\par
\noindent
In conclusion, we  performed  the
Ramsauer  model  parameterization  of experimental neutron total  scattering cross sections
from light and heavy nuclei. The parameters have been found to be same for light and heavy nuclei,
except for the radius parameters. We could reproduce the experimental neutron total cross sections by using twelve parameters
in the energy range of 20MeV to 550 MeV.
We proposed a new functional form for energy dependence and atomic mass dependence of the Ramsauer model
parameters. \\


\begin{references}
\bibitem{ads1}  C.  Rubbia  {\it  et.  al.,}  " Conceptual design of a fast
Neutron Operated High power Energy Amplifier",
CERN Rep,   CERN/AT/95-44 (ET), Geneva, 29 Sep. 1995.
\bibitem{ads2}IAEA-TEC-DOC-985,  Accelerator  Driven Systems ; Energy generation
and Transmutation of Waste Status Report, Nov  1997,  International  Atomic
Energy Agency, Vienna.
\bibitem{lawson} J. D. Lawson, Phil. Mag., {\bf 44}, 102 (1953).
\bibitem{finlay} R. W. Finlay, W. P. Abfalterer, G. Fink {\it et. al.,} Phys. Rev. {\bf C47}, 237 (1993).
\bibitem{dietrich1} F. S. Dietrich, W. P. Abfalterer, {\it et. al.,}
 $LANSCE-WNR$, Proc. Int. Conf. Nucl. data for Sci. and Tech.,
 Trieste, Italy, May 19-24, 1997, Vol.59, p.402, Italian Phys. Soc. (1997).
\bibitem{abfal}  W. P. Abfalterer, F. B. Bateman, {\it et. al.,} Phys. Rev {\bf C63}, 044608, (2001).
\bibitem{kd}A. J. Koning and J. -P. Delaroche, Nucl. Phys. {\bf A713}, 231 (2003).
\bibitem{peterson} J. M. Peterson, Phys. Rev. {\bf 125}, 955 (1962).
\bibitem{book}  A.  Bohr and B. Mottelson, Nuclear Structure, Vol. 1 P.166,
Benjamin, N.Y. (1969).
\bibitem{franco} V. Franco, Phys. Rev. {\bf B140}, 1501 (1965).
\bibitem{gould} C. R. Gould {\it et. al.,} Phys. Rev. Lett. {\bf 53}, 2371 (1986).
\bibitem{anderson} J. D. Anderson and S. M. Grimes,
Phys. Rev. {\bf C41}, 2904 (1990).
\bibitem{grimes1} S. M. Grimes, J. D. Anderson, R. W. Bauer
and V. A. Madsen, Nucl. Sci. and Engg. {\bf 130}, 340 (1998).
\bibitem{madsen} V. A. Madsen, J. D. Anderson, S. M. Grimes, V. R. Brown and P. M. Antony,
Phys. Rev. {\bf C56}, 365 (1997).
\bibitem{bauer} R. W. Bauer {\it et. al.,} Nucl. Sci. and Engg. {\bf  130},
348 (1998).
\bibitem{grimes2} S. M. Grimes, J. D. Anderson, R. W. Bauer
and V. A. Madsen, Nucl. Sci. and Engg. {\bf 134}, 77 (2000).
\bibitem{grimes3} S. M. Grimes, J. D. Anderson and R. W. Bauer,
Nucl. Sci. and Engg. {\bf 135}, 296 (2000).
\bibitem{dietrich2} F. S. Dietrich, J. D. Anderson and R. W. Bauer, Phys. Rev. {\bf C68}, 064608 (2003).
\bibitem{deb} P. K. Deb and K. Amos, Phys. Rev. {\bf 67}, 067602 (2003)~;~{\it ibid}  {\bf 69}, 064608 (2004)
\bibitem{surya1}S. S. V. Suryanarayan, Rajesh S. Gowda and S. Ganesan, arXiv: nucl-th/0409005. \bibitem{scat2} O. Bersillon, "SCAT2" Program, Note CEA-N-2227, Centre d'Etudes
 Nucleaires de Bruyeres-Chatel, Service de Physique et techniques Nucleaires, France (Oct. 1981).
\end{references}
\end{document}